\documentclass[prb,twocolumn]{revtex4}%

\usepackage{graphicx}
\usepackage{amsmath}
\usepackage{ulem}
\usepackage{color}

\newcommand{\be}{\begin{equation}}
\newcommand{\ee}{\end{equation}}
\newcommand{\bea}{\begin{eqnarray*}}
\newcommand{\eea}{\end{eqnarray*}}
\newcommand{\bean}{\begin{eqnarray}}
\newcommand{\eean}{\end{eqnarray}}

\begin{document}

\draft
\title{\bf Charge transport through the multiple end zigzag edge states of armchair graphene nanoribbons and heterojunctions}

\author{David M T Kuo}
%\DIFdelbegin \DIFdel{11
%}\DIFdelend

\address{Department of Electrical Engineering and Department of Physics, National Central
University, Chungli, 32001 Taiwan, China}

\date{\today}

\begin{abstract}
This comprehensive study investigates charge transport through the
multiple end zigzag edge states of finite-size armchair graphene
nanoribbons/boron nitride nanoribbons (n-AGNR/w-BNNR) junctions
under a longitudinal electric field, where n and w denote the
widths of the AGNRs and the BNNRs, respectively. In 13-atom wide
AGNR segments, the edge states exhibit a blue Stark shift in
response to the electric field, with only the long decay length
zigzag edge states showing significant interaction with the red
Stark shift subband states. Charge tunneling through such edge
states assisted by the subband states is elucidated in the spectra
of the transmission coefficient. In the 13-AGNR/6-BNNR
heterojunction, notable influences on the energy levels of the end
zigzag edge states of 13-AGNRs induced by BNNR segments are
observed. We demonstrate the modulation of these energy levels in
resonant tunneling situations, as depicted by bias-dependent
transmission coefficient spectra. Intriguing nonthermal broadening
of tunneling current shows a significant peak-to-valley ratio. Our
findings highlight the promising potential of n-AGNR/w-BNNR
heterojunctions with long decay length edge states in the realm of
GNR-based single electron transistors at room temperature.
\end{abstract}

\maketitle

\section{Introduction}
Graphene nanoribbons (GNRs) have been the subject of extensive
study since the groundbreaking discovery of two-dimensional
graphene in 2004 by Novoselov and Geim [\onlinecite{Novoselovks}].
Known for their semiconducting phases resulting from quantum
confinement, GNRs hold significant promise for next-generation
electronics [\onlinecite{JustinH}--\onlinecite{WangHM}]. Among
these, armchair GNRs (AGNRs) are particularly noteworthy due to
their tunable band gaps, which are inversely proportional to their
widths [\onlinecite{Cai}--\onlinecite{DJRizzo}]. Recent research
has focused on understanding the electronic properties of AGNRs
under transverse electric fields, revealing transitions from
semiconducting to semimetallic phases
[\onlinecite{ZhaoF}--\onlinecite{NikitaVT}]. These transitions
could be crucial for controlling plasmon propagation.

The terminal zigzag edge structures of AGNRs play a critical role
in field-effect transistors (FETs), directly interfacing with the
source and drain electrodes
[\onlinecite{LlinasJP},\onlinecite{JacobsePH}]. These structures
harbor topological states (TSs), the response of which to electric
fields remains poorly understood
[\onlinecite{TepliakovNV},\onlinecite{GolorM}]. Wider AGNRs
exhibit multiple terminal zigzag edge states in finite segments
[\onlinecite{SanchoMP},\onlinecite{Zdetsis}]. For instance, in
13-atom wide AGNR (13-AGNR) segments, each terminal may possess
two distinct zigzag edge states, one with an exponential decay
characteristic length and another with a longer characteristic
length. Although the current spectra of 13-AGNR tunneling FETs
have been experimentally reported, a systematic analysis of charge
tunneling through the topological edge states and the subband
states is lacking [\onlinecite{LlinasJP}]. Meanwhile, scientists
want to know how the electronic and transport properties of GNRs
are changed when GNRs are embedded into h-boron nitride sheets
[\onlinecite{NikitaVT}], which is a critical issue to realize
GNR-based electronic circuits
[\onlinecite{ChenLX}-\onlinecite{GengDC}].

This study aims to elucidate the electronic and transport
properties of multiple end zigzag edge states in finite-size AGNRs
and AGNR/boron nitride nanoribbon (BNNR) heterojunctions under
longitudinal electric fields. Figure 1 illustrates an
n-AGNR/w-BNNR heterojunction, where n and w denote the widths of
AGNR and BNNR, respectively. Given the challenges associated with
using density functional theory (DFT) to calculate the
transmission coefficient of GNR segments under electric fields
[\onlinecite{SonYW},\onlinecite{Matsuda}], we employ a
tight-binding model and Green's function technique to compute the
bias-dependent transmission coefficient
[\onlinecite{Kuo1}--\onlinecite{Kuo3}]. This approach enables us
to elucidate the charge transport through multiple zigzag edge
states in finite 13-AGNRs and 13-AGNR/6-BNNR heterojunctions.
Spectra of transmission coefficients reveal interference between
subband states and long characteristic length zigzag edge states
in the finite 13-AGNR structure. Furthermore, the on-off switch
behavior arising from resonant tunneling between the left and
right zigzag edge states of the 13-AGNR/6-BNNR heterojunction is
discerned in the transmission coefficient spectra. Intriguingly,
nonthermal broadening tunneling current exhibits a large
peak-to-valley ratio, which is highly useful for applications such
as single-electron transistors at room temperature.

\begin{figure}[h]
\centering
\includegraphics[trim=1.cm 0cm 1.cm 0cm,clip,angle=0,scale=0.3]{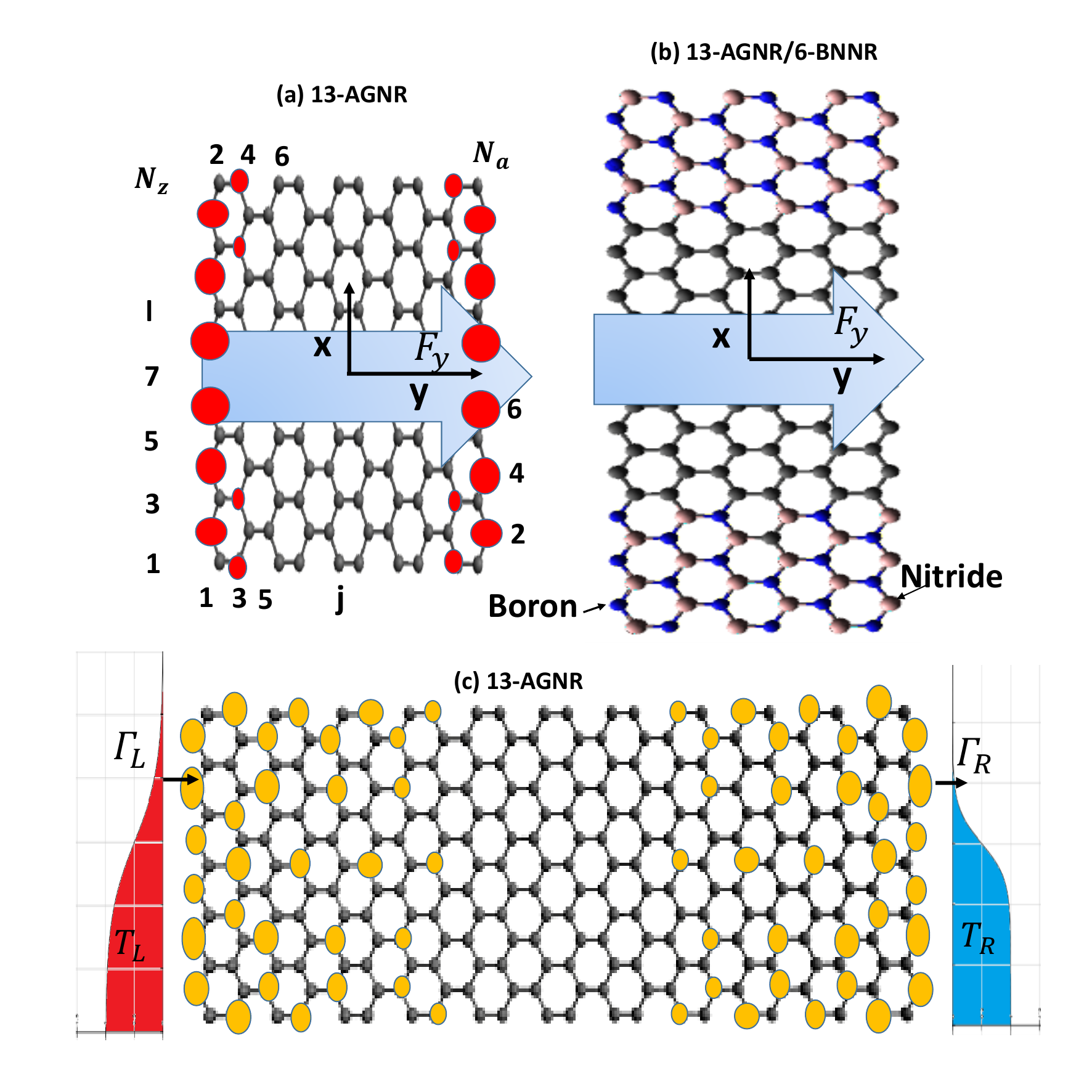}
\caption{(a) and (b) Schematic diagrams illustrating the armchair
graphene nanoribbon (13-AGNR) and 13-AGNR/6-BNNR heterojunction
under longitudinal electric fields, respectively. Schematic
diagram (c) depicts the line-contacting of zigzag-edge atoms in a
13-AGNR structure to electrodes. Symbols $\Gamma_{L}$ ($\Gamma_R$)
represent the electron tunneling rate between the left (right)
electrode and the leftmost (rightmost) atoms at the zigzag edges,
and $T_{L}$ ($T_{R}$) denotes the equilibrium temperature of the
left (right) electrode. The probability densities of the zigzag
edge states with short and long characteristic lengths in 13-AGNR
segments without electric fields are plotted in Figure 1(a) and
1(c), respectively. The radius of the circle represents the
magnitude of the probability density.}
\end{figure}
\section{Calculation Methodology}
To investigate the transport properties of 13-AGNR and
13-AGNR/6-BNNR heterojunctions connected to the electrodes, we
employ a combination of the tight-binding model and the Green's
function technique. The system Hamiltonian depicted in Fig.~1(c)
is written as $H=H_0+H_{GNR}$, where

\begin{small}
\begin{eqnarray}
H_0& = &\sum_{k} \epsilon_k a^{\dagger}_{k}a_{k}+
\sum_{k} \epsilon_k b^{\dagger}_{k}b_{k}\\
\nonumber &+&\sum_{\ell}\sum_{k}
V^L_{k,\ell,j}d^{\dagger}_{\ell,j}a_{k}
+\sum_{\ell}\sum_{k}V^R_{k,\ell,j}d^{\dagger}_{\ell,j}b_{k} + h.c.
\end{eqnarray}
\end{small}
The first two terms of Eq.~(1) describe the free electrons in the
left and right metallic electrodes. $a^{\dagger}_{k}$
($b^{\dagger}_{k}$) creates  an electron of with momentum $k$ and
energy $\epsilon_k$ in the left (right) electrode.
$V^L_{k,\ell,j=1}$ ($V^R_{k,\ell,j=N_a}$) describes the coupling
between the left (right) lead with its adjacent atom in the
$\ell$-th row.

\begin{small}
\begin{eqnarray}
H_{GNR}&= &\sum_{\ell,j} E_{\ell,j} d^{\dagger}_{\ell,j}d_{\ell,j}\\
\nonumber&-& \sum_{\ell,j}\sum_{\ell',j'} t_{(\ell,j),(\ell', j')}
d^{\dagger}_{\ell,j} d_{\ell',j'} + h.c,
\end{eqnarray}
\end{small}

Here, $E_{\ell,j}$ represents the on-site energy of the orbital in
the $\ell$-th row and $j$-th column. The operators
$d^{\dagger}_{\ell,j}$ and $d_{\ell,j}$ create and annihilate an
electron at the atom site denoted by $(\ell,j)$. The parameter
$t_{(\ell,j),(\ell', j')}$ characterizes the electron hopping
energy from site $(\ell',j')$ to site $(\ell,j)$. For the
tight-binding parameters of n-AGNR/w-BNNR, we assign $E_{B} =
2.329$ eV, $E_{N} = -2.499$ eV, and $E_{C} = 0$ eV to boron,
nitride, and carbon atoms, respectively. We neglect variations in
electron hopping strengths between different atoms for simplicity
[\onlinecite{GSSeal},\onlinecite{DingY}]. We set
$t_{(\ell,j),(\ell',j')} = t_{pp\pi} = 2.7$ eV for the
nearest-neighbor hopping strength. Additionally, the effect of
electric field $F_y$ is included by the electric potential $U = e
F_y y$ on $E_{\ell,j}$, where $F_y = V_y/L_a$, with $V_y$ as the
applied bias and $L_a$ as the length of the AGNR segment.

We calculate the bias-dependent transmission coefficient ${\cal
T}_{LR}(\varepsilon)$ using the formula: ${\cal
T}_{LR}(\varepsilon) =
4Tr[\Gamma_{L}(\varepsilon)G^{r}(\varepsilon)\Gamma_{R}(\varepsilon)G^{a}(\varepsilon)]$
[\onlinecite{Kuo1}], where $\Gamma_{L}(\varepsilon)$ and
$\Gamma_{R}(\varepsilon)$ denote the tunneling rate (in energy
units) at the left and right leads, respectively, and
$G^{r}(\varepsilon)$ and $G^{a}(\varepsilon)$ are the retarded and
advanced Green's functions of the GNRs, respectively. In terms of
tight-binding orbitals, $\Gamma_{\alpha}(\varepsilon)$ and Green's
functions are matrices. The expression for
$\Gamma_{L(R)}(\varepsilon)$ is derived from the imaginary part of
the self-energies, denoted as $\Sigma^r_{L(R)}(\varepsilon)$, and
is given by
$\Gamma_{L(R)}(\varepsilon)=-\text{Im}(\Sigma^r_{L(R)}(\varepsilon))=\pi\sum_k|V^{L(R)}_{k,\ell,j}|^2\delta(\varepsilon-\epsilon_k)$,
where $V^{L}_{k,\ell,j=1}$ and $V^{R}_{k,\ell,j=N_a}$ denote the
coupling strengths between the left and right metallic electrodes
with their adjacent atoms (refer to Fig. 1). In the context of
metals such as gold, where the density of states remains
approximately constant near the Fermi energy, the wide-band limit
serves as a suitable approximation. In this limit, the
energy-dependent tunneling rates, $\Gamma_{L(R)}(\varepsilon)$,
are replaced by constant matrices denoted as
$\Gamma_{L(R)}$[\onlinecite{Kuo1}]. However, it's important to
note that in some cases, only the diagonal entries of
$\Gamma_{L(R)}$ are non-zero. Additionally, if one adopts the
alternative definition of
$\Gamma_{L(R)}(\varepsilon)=-i[\Sigma^r_{L(R)}(\varepsilon)-\Sigma^a_{L(R)}(\varepsilon)]$,
the factor "4" in the transmission coefficient is reset to one
[\onlinecite{SunQF}--\onlinecite{YuXX}]. Notably, $\Gamma_{L(R)}$
calculated within the wide-band limit can effectively replicate
the transmission coefficient curves of graphene nanoribbon
superlattices obtained using the surface Green's function method,
as demonstrated in references [\onlinecite{NingXU}] and
[\onlinecite{YuXX}].

\section{Results and discussion}
Before examining the effects of electric fields on the electronic
and transport properties of end zigzag edge states of 13-AGNR and
13-AGNR/6-BNNR structures, we first present the electronic
structures of these two configurations in Fig. 2(a) and Fig. 2(b),
respectively. Both structures maintain semiconductor properties
with direct band gaps. Specifically, the 13-AGNR and
13-AGNR/6-BNNR exhibit band gaps of $0.714$ eV and $0.722$ eV,
respectively, consistent with values calculated by the DFT method
[\onlinecite{DingY}]. It's noteworthy that the electronic
structures of 13-AGNR/w-BNNR remain unchanged when $w \geq 6$.
This suggests that a 6-BNNR width is sufficient to model 13-AGNR
embedded within an h-BN sheet. Although the impact of BNNR on the
band gaps of 13-AGNRs is not significant, we will demonstrate
later that BNNR structures have a notable effect on the end zigzag
edge states of 13-AGNRs.

\subsection{13-AGNR Segments under an Electric Field}

To elucidate the end zigzag edge states of 13-AGNRs, we analyze
the energy spectra of 13-AGNR segments under an electric field.
Figures 2(c) and 2(d) illustrate the energy levels of 13-AGNR
segments with $N_a = 64$ ($L_a = 6.67$ nm) and $N_a = 200$ ($L_a =
21.16$ nm) as functions of $V_y$. In the absence of an electric
field, four energy levels exist between $E_C$ and $E_V$,
representing the subband states as depicted in Fig. 2(c). These
levels are denoted as $\Sigma_{C,S}$, $\Sigma_{C,L}$,
$\Sigma_{V,S}$, and $\Sigma_{V,L}$. Here, $C$ and $V$ distinguish
the energy levels above or below the Fermi energy, while $S$ and
$L$ represent short and long decay characteristic lengths,
respectively. These levels arise from the formation of bonding and
antibonding states between the left and right TSs ($\Psi_{L,S(L)}$
and $\Psi_{R,S(L)}$), as noted in ref [\onlinecite{SanchoMP}]. For
$N_a = 64$, $\Sigma_{C(V),L} = \pm 7.13$ meV and $\Sigma_{C(V),S}
= 0$. For $N_a = 200$, $\Sigma_{C(V),L} = \pm 7 \mu$eV and
$\Sigma_{C(V),S} = 0$. When considering $N_a = 200$, $E_C$ and
$E_V$ are $0.3668$ eV and $-0.3668$ eV, respectively, which
closely correspond to the minimum conduction subband and the
maximum valence subband of infinitely long 13-AGNRs shown in Fig.
2(a). Notably, in Fig. 2(c) and 2(d), we observe a blue Stark
shift of TSs and a red Stark shift of the subband states.
Consequently, the energy levels of end zigzag edge states cross
the subband states. With increasing $N_a$, the applied bias
inducing such a crossing is shifted toward lower bias, as observed
in Fig. 2(d). Note that the Stark shift of energy levels for
graphene nanoribbon (GNR) segments (or graphene quantum dots)
responds sensitively to the applied electric fields, with
directionality playing a crucial role
[\onlinecite{ChenRB},\onlinecite{PedersenTG}]. Particularly, when
transverse electric fields ($F_x$) are introduced to armchair
graphene nanoribbons (AGNRs), we observe an intriguing
semiconductor-to-semimetal transition in the electronic structures
[\onlinecite{ZhaoF},\onlinecite{Pizzochero}].

\begin{figure}[h]
\centering
\includegraphics[angle=0,scale=0.3]{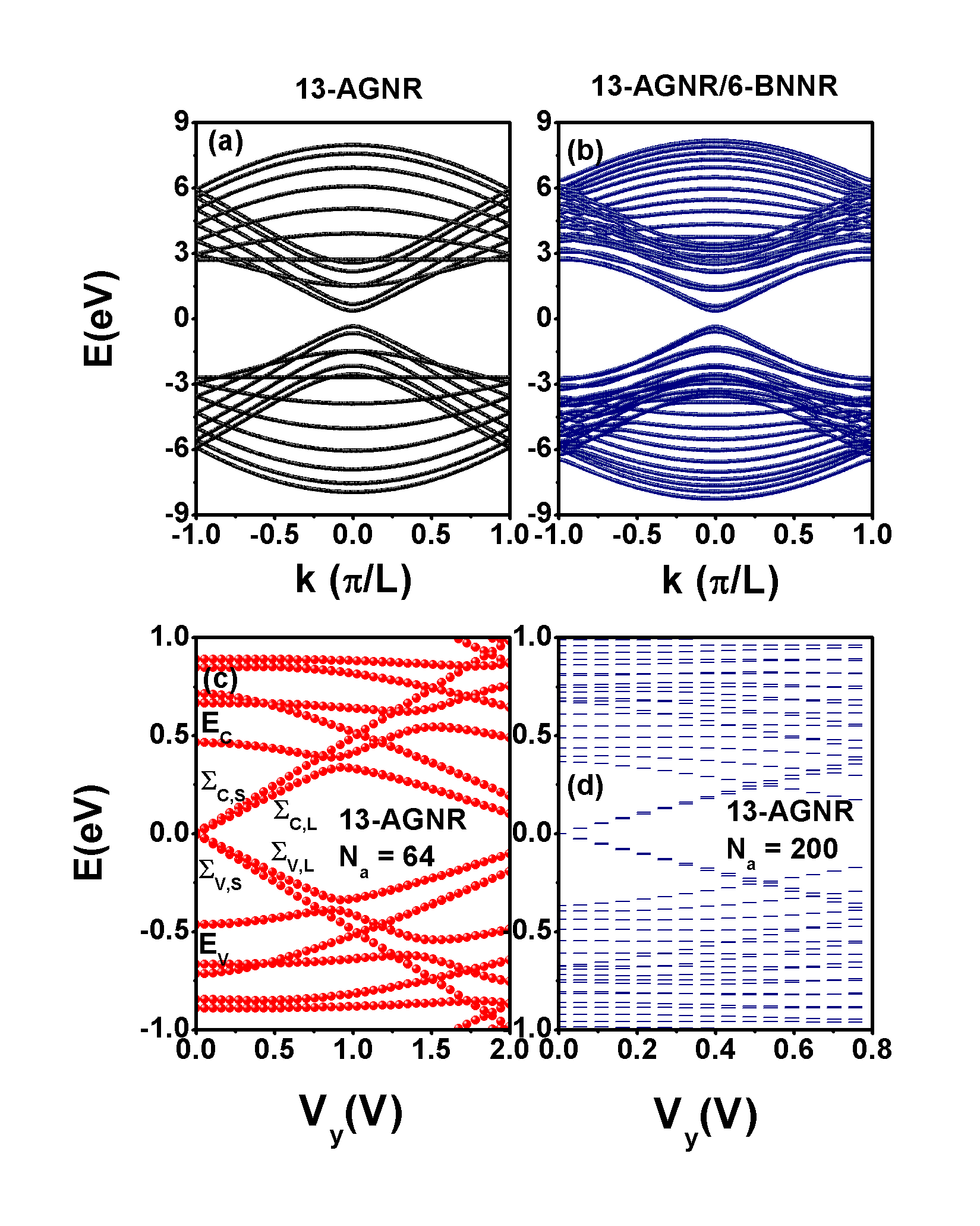}
\caption{Electronic structures of 13-AGNR and 13-AGNR/6-BNNR
structures. (a) 13-AGNR structure and (b) 13-AGNR/6-BNNR
structure. Energy levels of 13-AGNRs as functions of applied
voltage $V_y$ for two different $N_a$ values. (c) $N_a = 64$ ($L_a
= 6.67$ nm) and (d) $N_a = 200$ ($L_a = 21.16$ nm). Note that the
energy levels are plotted with respect to the Fermi energy $E_F$,
which is set as zero throughout this article.}
\end{figure}

To further comprehend the charge transport through these
topological states ($\Psi_{L(R),S}$ and $\Psi_{L(R),L}$), we plot
the transmission coefficient ${\cal T}_{LR}(\varepsilon)$ of a
13-AGNR segment with $N_a = 64$ for various $V_y$ values in Fig.
3(a)-3(c). As depicted in Fig. 3(a), the peak labeled by
$\Sigma_0$ arises from $\Sigma_{C,L}$ and $\Sigma_{V,L}$, rather
than $\Sigma_{C,S}$ and $\Sigma_{V,S}$, as their wave functions
($\Psi_{L,S}$ and $\Psi_{R,S}$) are highly localized on the end
zigzag edges (see Fig. 1(a)). With the application of bias voltage
$V_y$ in Fig. 3(b), the probability for charge tunneling through
these topological states diminishes significantly due to the
off-resonance energy levels of $\Psi_{L,L}$ and $\Psi_{R,L}$. Upon
reaching $V_y = 0.918$ V, two peaks labeled $\varepsilon_{C(V),B}$
and $\varepsilon_{C(V),A}$ emerge due to the interference between
the end zigzag edge states $\Sigma_{C(V),L}$ and the subband
states $E_{C(V)}$, as seen in Fig. 3(c). The peak at
$\varepsilon_{C(V),B}$ can be interpreted as charge tunneling
through the long decay length edge state under the subband states
($E_{C(V)}$) assisted procedure. Fig. 3(d) illustrates the contact
effect on this interference phenomenon. As $\Gamma_t$ decreases,
the resolution between the peaks ($\varepsilon_{C,B}$ and
$\varepsilon_{C,A}$) becomes more distinct. For $\Gamma_t = 90$
meV, the interference pattern resembles Fano interference. The
probability distributions of the peaks labeled $\varepsilon_{C,B}$
and $\varepsilon_{C,A}$ are shown in Fig. 3(e) and 3(f),
respectively. It is evident that $\varepsilon_{C(V),B}$ and
$\varepsilon_{C(V),A}$ correspond to constructive and destructive
interferences between the wave function of the conduction
(valence) subband state and the wave function of the right (left)
edge state with a long characteristic length ($\Psi_{R(L),L}$).
Although many theoretical efforts predict the existence of
topological states of GNRs
[\onlinecite{DRizzo}-\onlinecite{DJRizzo}], revealing the quantum
interference between the TSs and the subband states based on the
tunneling spectra of GNR-based electronic devices remains a
challenge [\onlinecite{LlinasJP},\onlinecite{JacobsePH}].

\begin{figure}[h]
\centering
\includegraphics[angle=0,scale=0.3]{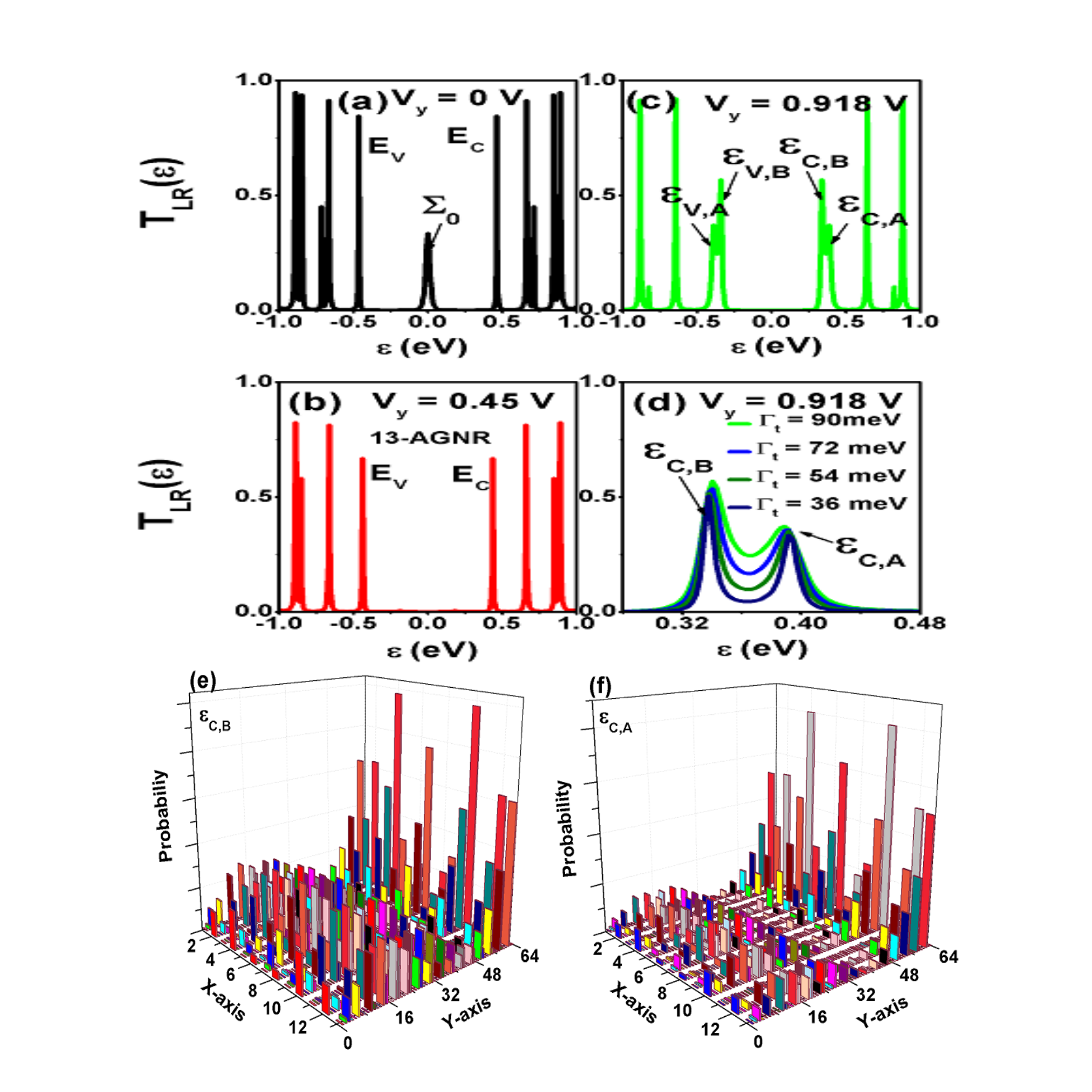}
\caption{Transmission coefficients ${\cal T}_{LR}(\varepsilon)$ of
13-AGNRs with $N_a = 64$ for various $V_y$ values at $\Gamma_t =
0.09$ eV. Panels (a), (b), and (c) correspond to $V_y = 0$, $V_y =
0.45$ V, and $V_y = 0.918$ V, respectively. Panel (d) shows the
${\cal T}_{LR}(\varepsilon)$ for various $\Gamma_t$ values at $V_y
= 0.918$ V. Panels (e) and (f) depict probability densities of
$\varepsilon_{C,B} = 0.337$ eV and $\varepsilon_{C,A} = 0.393$ eV
in a 13-AGNR segment with $N_a = 64$ and $V_y = 0.918$ V.}
\end{figure}

\subsection{13-AGNR/6-BNNR Segments under an Electric Field}

Considering the influence of h-BN on the electronic and transport
properties of 13-AGNRs, we present the energy levels of
13-AGNR/6-BNNR structures as functions of applied $V_y$ for
various $N_a$ values in Fig. 4(a)-(c). In the gap region of the
13-AGNR/6-BNNR structure (refer to Fig. 2(b)), there are still
four energy levels present ($\Sigma_{C,S}$, $\Sigma_{C,L}$,
$\Sigma_{V,S}$, and $\Sigma_{V,L}$). Unlike the multiple end
zigzag edge states of 13-AGNRs, the multiple zigzag edge states of
finite 13-AGNR/6-BNNR heterojunctions are insensitive to
variations in $N_a$, remaining nearly fixed at specific energy
levels. For instance, at $N_a = 44$, we have $\Sigma_{C,S} = 76.4$
meV ($\Sigma_{V,S} = -81.7$ meV) and $\Sigma_{C,L} = 0.253$ eV
($\Sigma_{V,L} = -0.268$ eV). Similarly, at $N_a = 64$, we observe
$\Sigma_{C,S} = 76.4$ meV ($\Sigma_{V,S} = -81.7$ meV) and
$\Sigma_{C,L} = 0.249$ eV ($\Sigma_{V,L} = -0.263$ eV). This
indicates that $\Sigma_{C,S}$, $\Sigma_{C,L}$, $\Sigma_{V,S}$, and
$\Sigma_{V,L}$ are not determined by the wave function overlaps
between the left and right TSs. In finite 13-AGNR/6-BNNR
heterojunctions, BNNRs have the left boron atom zigzag edge
terminal and the right nitride atom zigzag edge terminal (see Fig.
1(b)). The energy levels of $\Sigma_{C,S}$, $\Sigma_{C,L}$,
$\Sigma_{V,S}$, and $\Sigma_{V,L}$ are significantly influenced by
the energy levels of boron and nitride atoms.

Figures 4(a)-4(c) indicate that the multiple zigzag edge states of
13-AGNR/6-BNNR exhibit a linear function of the electric field. In
the high bias region ($V_y > 1$ V), the interaction spectra
resulting from the multiple end zigzag edge states and the subband
states can be reproduced, akin to the interference scenario
observed in 13-AGNRs (Fig. 2(c) and Fig. 2(d)). Before calculating
the transmission coefficient of 13-AGNR/6-BNNR heterojunction, we
plot the probability densities of $\Sigma_{C,L}$ and
$\Sigma_{V,L}$ at $V_y = 0$ and $N_a = 44$ in Fig. 4(d) and 4(e),
respectively. From the probability distributions of $\Sigma_{C,L}$
and $\Sigma_{V,L}$, it is evident that charges are confined within
the 13-AGNR segment, demonstrating that BNNRs act as potential
barriers. Additionally, $\Sigma_{C,L}$ ($\Sigma_{V,L}$) is
determined by the unoccupied carbon-nitride antibonding state
(occupied carbon-boron bonding
state)[\onlinecite{PrunedaJM},\onlinecite{JungJ}].

The maximum probability of $\Sigma_{C,L}$ ($\Sigma_{V,L}$) occurs
at sites labeled by $\ell = 10$ and $\ell = 16$ at $j = 1$ ($\ell
= 10$ and $\ell = 16$ at $j = 44$). The probability distribution
of $\Sigma_{C,L}$ ($\Sigma_{V,L}$) along the $y$ direction
exhibits a long decay length, spanning several benzene sizes, yet
its probability density near the right (left) electrode sites is
minimal. This indicates that the wave function of $\Sigma_{C,L}$
($\Sigma_{V,L}$) is weakly coupled to the right (left) electrode.
The asymmetrical coupling to the left and right electrodes for the
wave function of $\Sigma_{C,L}$ ($\Sigma_{V,L}$) hinders charge
transport through this energy level.

\begin{figure}[h]
\centering
\includegraphics[angle=0,scale=0.3]{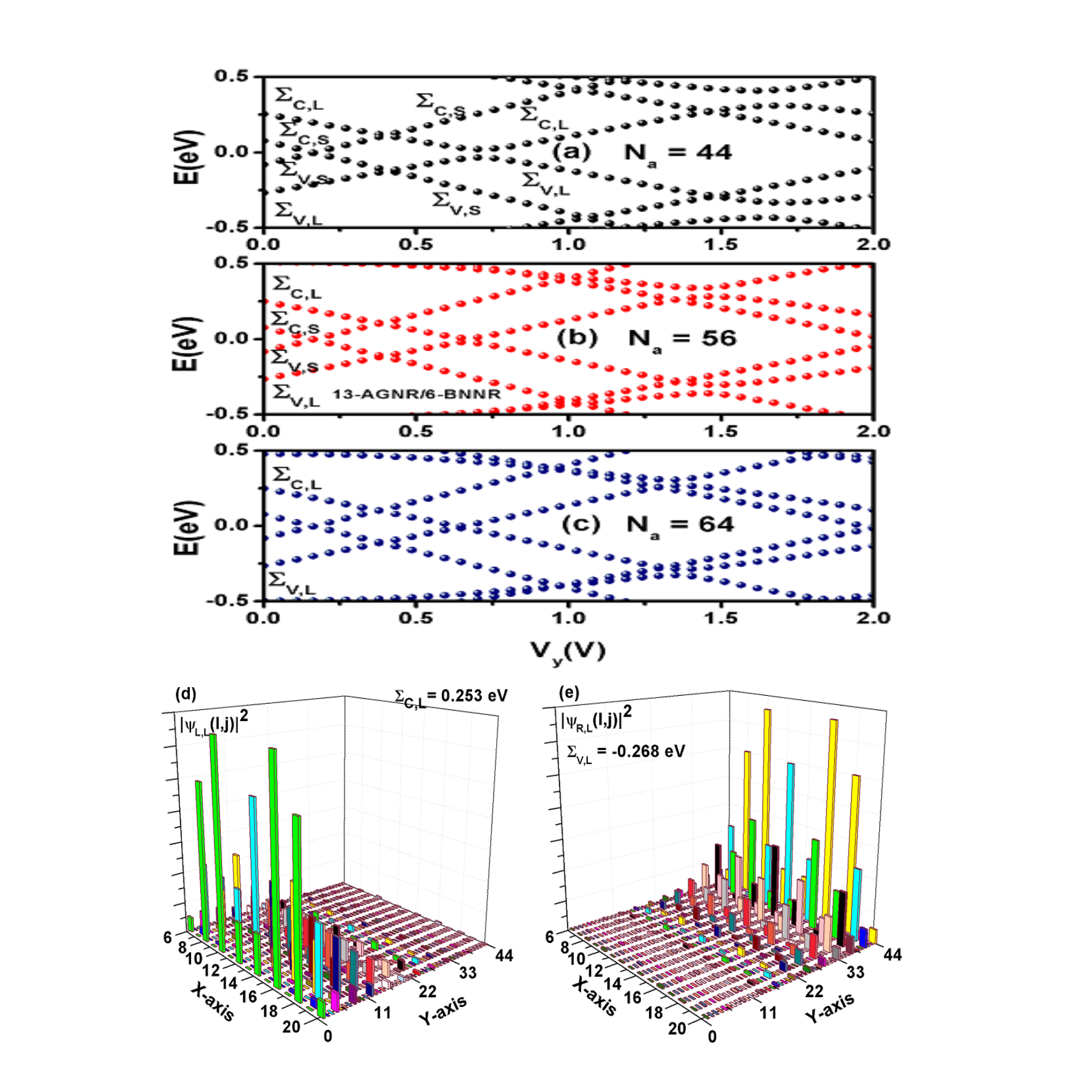}
\caption{Energy levels of 13-AGNR/6-BNNR heterojunction as
functions of $V_y$ for various $N_a$. Panels (a), (b), and (c)
correspond to $N_a = 44$ ($L_a = 4.54$ nm), $N_a = 56$ ($L_a =
5.82$ nm), and $N_a = 64$ ($L_a = 6.67$ nm), respectively.
Probability densities of 13-AGNR/6-BNNR heterojunction with $N_a =
44$ ($L_a = 4.54$ nm) at $V_y = 0$. Panels (d) and (e) depict the
probability densities of $\Sigma_{C,L}$ and $\Sigma_{V,L}$,
respectively.}
\end{figure}

We now present the calculated transmission coefficient of the
13-AGNR/6-BNNR heterojunction with $N_a = 44$ for various applied
bias $V_y$ values at $\Gamma_t = 90$ meV in Fig. 5. In Fig. 5(a)
and 5(b), the charge transport through $\Sigma_{C,L}$ and
$\Sigma_{V,L}$ is hindered due to the very weak coupling strength
between the electrodes and the zigzag edge states. However, a
significant peak of $\Sigma_0$ occurs for $\varepsilon$ near the
Fermi energy at $V_y = 0.702$ V in Fig. 5(c). For $V_y = 1.02$ V
and $V_y = 1.458$ V, the peaks for charge transport through
$\Sigma_{C,L}$ and $\Sigma_{V,L}$ are shifted away from the Fermi
energy, as shown in Fig. 5(d) and 5(e). A notable enhancement of
charge transport through $\Sigma_{C,L}$ and $\Sigma_{V,L}$ is
observed for $V_y = 1.62$ V, which is attributed to the assistance
from subband states.

The transmission coefficient of $\Sigma_0$ shown in Fig. 3(c) is
attributed to the alignment between $\Sigma_{C,L}$ and
$\Sigma_{V,L}$. We plot the probability distribution of $\Sigma_0
= 0$ in Fig. 5(g). The symmetrical probability distribution
explains the alignment of $\Sigma_{C,L}$ and $\Sigma_{V,L}$. As
these two levels approach alignment, their wave functions
$\Psi_{C,L}(\ell,j)$ and $\Psi_{V,R}(\ell,j)$ generate
constructive and destructive superpositions, forming the two
splitting peaks. To resolve these two peaks, we present the
transmission coefficient ${\cal T}_{LR}(\varepsilon)$ for various
$\Gamma_t$ values in Fig. 5(h). The results in Fig. 5(h) indicate
that the transmission coefficient is affected not only by the
applied voltage but also by the contact property between the
electrodes and the 13-AGNR/6-BNNR heterojunctions. For weak
$\Gamma_t = 36$ meV, we observe two separated peaks.

\begin{figure}[h]
\centering
\includegraphics[angle=0,scale=0.3]{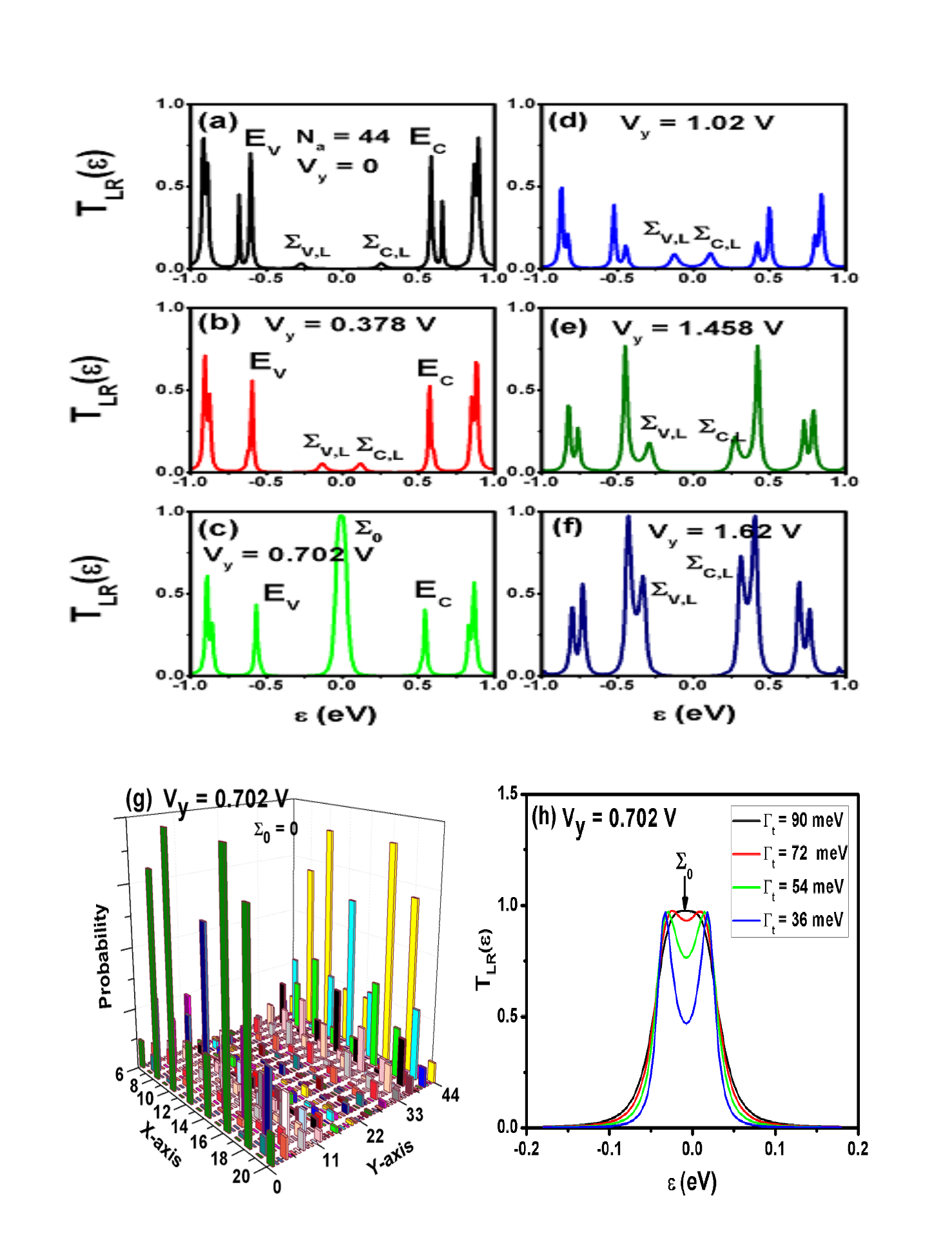}
\caption{Transmission coefficient ${\cal T}_{LR}(\varepsilon)$ of
the 13-AGNR/6-BNNR heterojunction with $N_a = 44$ for various
$V_y$ values at $\Gamma_t = 90$ meV. Panels (a) through (f)
correspond to $V_y$ values of $0$, $0.378$ V, $0.702$ V, $1.02$ V,
$1.458$ V, and $1.62$ V, respectively. Panel (g) displays the
probability distribution of $\Sigma_{0} = 0$ at $V_y = 0.702$ V.
Panel (h) shows the transmission coefficient ${\cal
T}_{LR}(\varepsilon)$ as functions of $\varepsilon$ for various
$\Gamma_t$ values at $V_y = 0.702$ V.}
\end{figure}

\subsection{Tunneling Current through the Zigzag Edge States of 13-AGNR/6-BNNR Heterojunctions }
In this subsection, we investigate the tunneling current through
the end zigzag edge states with long decay length, utilizing the
effective 2-site Hubbard model [\onlinecite{Mangnus}]. The
transmission coefficient curves depicted in Figure 3(d) and Figure
5(h) can be faithfully replicated by the effective 2-site Hubbard
model, as the zigzag edge states are well-separated from the
subband states under low applied bias, and their wave functions
are localized [\onlinecite{Kuo2}]. The tunneling current is
determined by the expression:

\begin{equation}
J=\frac{-2e}{h}\int d\varepsilon~ {\cal
T}_{2-site}(\varepsilon)[f_L(\varepsilon)-f_R(\varepsilon)]
\end{equation}
where $f_{L(R)}(\varepsilon)$ denotes the Fermi distribution
function of the left (right) electrode with the chemical potential
$\mu_{L(R)} = E_F \mp eV_y/2$. The bias-dependent transmission
coefficient, ${\cal T}_{2-site}(\varepsilon)$, includes
one-particle, two-particle, and three-particle correlation
functions in the Coulomb blockade region, computed
self-consistently [\onlinecite{DavidK}]. Furthermore, the
bias-independent intra-site ($U_0$) and inter-site ($U_1$) Coulomb
interactions are computed using
$U_{n}=\frac{1}{4\pi\epsilon_0}\sum_{i,j}|\Psi_{C(V),L}(\textbf{r}_i)|^2|\Psi_{C(V),L}(\textbf{r}_j)|^2\frac{1}{|\textbf{r}_i-\textbf{r}_j|}$
with the dielectric constant $\epsilon_0 = 4$, and $U_{cc} = 4$ eV
at $i = j$. $U_{cc}$ arises from the two-electron occupation in
each $p_z$ orbital. In Fig. 6, the non-interaction curve
corresponds to the case depicted in Fig. 4(c). The inset of Fig. 6
illustrates charge transport in the Coulomb blockade region, where
$\varepsilon_{C,L} = \Sigma_{C,L}-\eta~eV_y+i\Gamma_{e,L}$ and
$\varepsilon_{V,L} = \Sigma_{V,L}+\eta~eV_y+i\Gamma_{e,R}$. Here,
$\eta~eV_y = 0.336~eV_y$ represents the orbital offset terms
induced by the applied voltage ($V_y$), and
$\Gamma_{e,L}=\Gamma_{e,R}=\Gamma_{e,t}$ represents the effective
tunneling rate of end zigzag states in the 2-site model,
determined by $\Gamma_t$ and the wave functions of the edge states
$\Psi_{C(V),L}$.

The first peak of the red curve, calculated at temperature $T =
300K$ and labeled $\varepsilon_1$, corresponds to the charge
transfer from $\varepsilon_{V,L} + U_0$ to $\varepsilon_{C,L}$
when these two energy levels align. The second peak,
$\varepsilon_2$, corresponds to electron transfer between
$\varepsilon_{V,L}$ and $\varepsilon_{C,L}$. However, the
tunneling current for such a procedure is significantly suppressed
due to electron Coulomb interactions. The blue curve with markers
is calculated at $T = 480K$. Distinguishing between the red and
blue curves in the entire applied voltage region is challenging,
illustrating the nonthermal broadening effect of the tunneling
current. Additionally, the peak-to-valley ratio reaches 8.7. The
results presented in Fig. 6 demonstrate that the edge states with
long decay length act as effective single charge filters at room
temperature, suggesting their potential application in
single-electron transistors at room temperature
[\onlinecite{LeeYM}]. Traditional single electron transistors
formed by a single quantum dot exhibit temperature-dependent
tunneling current spectra
[\onlinecite{LeobandungE}--\onlinecite{PostmaHWC}], where the
peak-to-valley ratio of tunneling current tends to decrease with
increasing temperature. Our study demonstrates that the tunneling
current through the zigzag edge states of 13-AGNR/6-BNNR
heterojunctions maintains temperature stability, a significant
advantage for practical applications.

\begin{figure}[h]
\centering
\includegraphics[angle=0,scale=0.3]{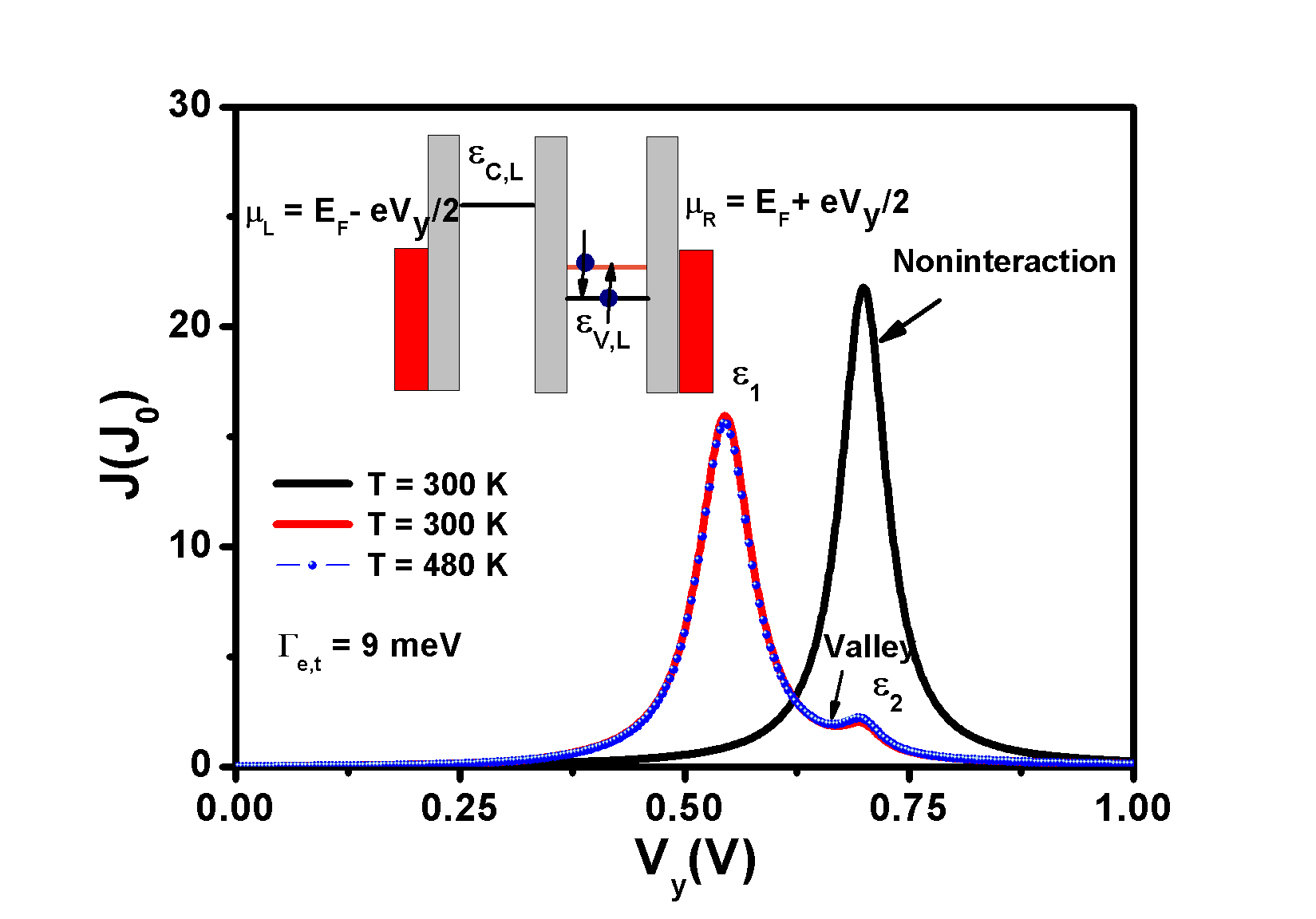}
\caption{Tunneling current through the end zigzag edge state with
a long decay length of the 13-AGNR/6-BNNR heterojunction with $N_a
= 64$. Energy levels are defined as $\varepsilon_{C,L} =
\Sigma_{C,L}-0.366eV_y+i\Gamma_{e,L}$ and $\varepsilon_{V,L} =
\Sigma_{V,L}+0.366eV_y+i\Gamma_{e,R}$. Other physical parameters
include $\Sigma_{C,L} = 0.249$ eV, $\Sigma_{V,L} = -0.263$ eV,
intra-site Coulomb interaction $U_0 = 155$ meV, inter-site Coulomb
interaction $U_1=42$ meV, and inter-site electron hopping strength
$t_{LR} = 7.13$ meV. We have
$\Gamma_{e,L}=\Gamma_{e,R}=\Gamma_{e,t} = 9$ meV, $E_F = 0$, and
$J_0 = 0.773$ nA.}
\end{figure}

\section{Conclusion}
In this study, we investigated the charge transport properties
through the multiple end zigzag edge states of both 13-AGNR and
13-AGNR/6-BNNR heterojunctions under longitudinal electric fields,
employing the tight-binding model and Green's function technique.
For 13-AGNRs, we elucidated the distinct blue Stark shift
behaviors exhibited by the zigzag edge states with short and long
decay lengths in response to electric fields. We found that the
latter can significantly interact with the subband states, thereby
unveiling the mechanism of charge transport through the long decay
length zigzag edge states assisted by the subband states, as
evidenced by the spectra of the transmission coefficient.

Concerning 13-AGNR/6-BNNR heterojunctions, we observed that the
energy levels of the end zigzag edge states of 13-AGNRs are
notably influenced by the presence of BNNRs rather than their
lengths. Even in long 13-AGNR/6-BNNR segments, $\Sigma_{C,L}$
exhibits a significant orbital offset from $\Sigma_{V,L}$. We
further revealed a remarkable resonant tunneling process for
charge transfer between $\Sigma_{C,L}$ and $\Sigma_{V,L}$ through
the bias-dependent transmission coefficient analysis.

Moreover, employing a 2-site Hubbard model, we analyzed the
tunneling current through the end zigzag edge states of
13-AGNR/6-BNNR heterojunctions within the Coulomb blockade region.
Our results demonstrate that these edge states function
effectively as single charge filters at room temperature,
exhibiting a high peak-to-valley ratio of tunneling current with a
nonthermal broadening effect, as illustrated in Fig. 6. This
unique characteristic suggests promising applications of
13-AGNR/6-BNNR heterojunctions in single electron transistors
operating at room temperature.

%\begin{flushleft}

%\end{flushleft}
{}
%\begin{flushleft}
\textbf{Conflicts of interest}\\
There are no conflicts to declare

\textbf{Data availability}\\

The data presented in this study are available upon reasonable
request.

{\bf Acknowledgments}\\
This work was supported by the National Science and Technology
Council, Taiwan under Contract No. MOST 107-2112-M-008-023MY2.
%\end{flushleft}

\mbox{}\\
E-mail address: mtkuo@ee.ncu.edu.tw\\
%E-mail address: yiachang@gate.sinica.edu.tw\\

%\appendix
%\numberwithin{figure}{section}
%\section{Electronic band structures, and ZT optimization of textured GNRs}
 \numberwithin{figure}{section} \numberwithin{equation}{section}

%\section{Appendix}~\Roman{section}
\setcounter{section}{0}
 %\section{Appendix}
\setcounter{equation}{0} % reset counter

%\section{}
%\subsection{Derivation of the tunneling current formula using Dyson's equations\label{App:TC_l} }
\mbox{}\\

%\appendix
%\numberwithin{figure}{section}
%\section{Electronic band structures}

%\numberwithin{equation}{section}

%{\bf Data Availability Statements}\\

%The data that supports the findings of this study are available
%within the article.

\end{document}